\documentstyle[amssymb]{article}

\input{tcilatex}

\begin{document}

\begin{center}
\ {\large THE HIDDEN SU(2) SUBSTRUCTURE OF QUARKS}

{\large AND THE KOBAIASHI-MASKAWA MATRIX}
\end{center}

{\large \bigskip }

\begin{center}
\noindent M\'{a}rio Everaldo de Souza

\noindent {\it Department of Physics, Universidade Federal de Sergipe, }

{\it Campus Universit\'{a}rio, 49100-000 S\~{a}o Crist\'{o}v\~{a}o, SE,
Brazil}
\end{center}

\bigskip

\parbox{4.3in}
{\noindent {\small Abstract. - It is proposed that each quark is composed of 
two prequarks, called primons. This composition generates a new hypercharge 
and a new kind of color called supercolor. Such quark composition explains 
the Kobaiashi-Maskawa matrix elements in terms of selection rules based on 
the new hypercharge. It is also shown that there should exist three Higgs 
bosons involved in the generation of quark masses.}}

\bigskip

\noindent {\large Introduction}

In recent years there have been attempts to go beyond the standard model in
particle physics. Going beyond the standard model means breaking up quarks
and/or leptons. Since the mass of the electron is already too small for a
particle with a very small radius which is smaller than 0.01 F, we can
consider it as being an elementary particle. Thus, in this article we
propose that leptons are elementary and that quarks are composite. Actually,
quark composition is an old idea$^{(1,2,3,4)}$. In order to distinguish the
quark composition proposed in this work from other models of the literature
we will name these prequarks with a different name. We call them primons, a
word derived from the Latin word primus which means first. Some preliminary
ideas on this subject have lately been published by the author$^{(5,6,7,8)}$%
.\bigskip

\noindent {\large The Number of Quarks and Primons, and the Supercolors}

Let us develop some preliminary ideas which will help us towards the
understanding of quark composition. Since a baryon is composed of three
quarks it is reasonable to consider that a quark is composed of two primons.
The new interaction between them exists by means of the exchange of new
bosons to be found.

In order to reproduce the spectrum of 6 quarks and their colors we need 4
primons in 3 supercolor states. Each color is formed by the two supercolors
of two different primons that form a particular quark. Therefore, the
symmetry group associated to the supercolor filed is SU(2). As to the
electric charge, one has charge (+5/6)e and any other one has charge (-1/6)e.

Taking into account the above considerations on colors and electric charges
of quarks we have the following tables for primons (Tables 1, 2, and 3).
According to Table 1 the maximum number of quarks is six. There should exist
similar tables for the corresponding antiparticles.

\bigskip

\begin{center}
\begin{tabular}{|ccc||ccc|ccc|ccc|}
\hline
&  &  &  &  &  &  &  &  &  &  &  \\ 
&  &  &  & $\alpha $ &  &  & $\beta $ &  &  & $\gamma $ &  \\ 
&  &  &  &  &  &  &  &  &  &  &  \\ \hline\hline
&  &  &  &  &  &  &  &  &  &  &  \\ 
& $\alpha $ &  &  &  &  &  & blue &  &  & green &  \\ 
&  &  &  &  &  &  &  &  &  &  &  \\ \hline
&  &  &  &  &  &  &  &  &  &  &  \\ 
& $\beta $ &  &  & blue &  &  &  &  &  & red &  \\ 
&  &  &  &  &  &  &  &  &  &  &  \\ \hline
&  &  &  &  &  &  &  &  &  &  &  \\ 
& $\gamma $ &  &  & green &  &  & red &  &  &  &  \\ 
&  &  &  &  &  &  &  &  &  &  &  \\ \hline
\end{tabular}
\end{center}

\vskip .1in

\begin{center}
\parbox{3in}
{Table 1. Generation of colors from supercolors}
\end{center}

\bigskip

\begin{center}
\begin{tabular}{|ccc||ccc|}
\hline
& superflavor &  &  & charge &  \\ \hline\hline
&  &  &  &  &  \\ 
& $p_{1}$ &  &  & $+\frac{5}{6}$ &  \\ 
&  &  &  &  &  \\ \hline
&  &  &  &  &  \\ 
& $p_{2}$ &  &  & $-\frac{1}{6}$ &  \\ 
&  &  &  &  &  \\ \hline
&  &  &  &  &  \\ 
& $p_{3}$ &  &  & $-\frac{1}{6}$ &  \\ 
&  &  &  &  &  \\ \hline
&  &  &  &  &  \\ 
& $p_{4}$ &  &  & $-\frac{1}{6}$ &  \\ 
&  &  &  &  &  \\ \hline
\end{tabular}
\end{center}

\vskip .1in

\begin{center}
\parbox{3in}
{Table 2. Electric charges of primons}
\end{center}

\bigskip

\vspace*{0.5in}

\bigskip

\begin{center}
\begin{tabular}{|ccc||ccc|ccc|ccc|ccc|}
\hline
&  &  &  &  &  &  &  &  &  &  &  &  &  &  \\ 
&  &  &  & $p_{1}$ &  &  & $p_{2}$ &  &  & $p_{3}$ &  &  & $p_{4}$ &  \\ 
&  &  &  &  &  &  &  &  &  &  &  &  &  &  \\ \hline\hline
&  &  &  &  &  &  &  &  &  &  &  &  &  &  \\ 
& $p_{1}$ &  &  &  &  &  & u &  &  & c &  &  & t &  \\ 
&  &  &  &  &  &  &  &  &  &  &  &  &  &  \\ \hline
&  &  &  &  &  &  &  &  &  &  &  &  &  &  \\ 
& $p_{2}$ &  &  & u &  &  &  &  &  & d &  &  & s &  \\ 
&  &  &  &  &  &  &  &  &  &  &  &  &  &  \\ \hline
&  &  &  &  &  &  &  &  &  &  &  &  &  &  \\ 
& $p_{3}$ &  &  & c &  &  & d &  &  &  &  &  & b &  \\ 
&  &  &  &  &  &  &  &  &  &  &  &  &  &  \\ \hline
&  &  &  &  &  &  &  &  &  &  &  &  &  &  \\ 
& $p_{4}$ &  &  & t &  &  & s &  &  & b &  &  &  &  \\ 
&  &  &  &  &  &  &  &  &  &  &  &  &  &  \\ \hline
\end{tabular}
\end{center}

\vskip .2in

\begin{center}
\parbox{3in}
{Table 3. Composition of quark flavors}
\end{center}

\bigskip

\noindent {\large The New Hypercharge and the new SU(2)}

In order to find the new hypercharge let us recall the relation between
electric charge and baryon number in quarks. Quark charges $2/3$ and $-1/3$
are symmetric about $1/6$ (Figure1) and, since $2/3-(-1/3)=1=2\times (1/2)$,
we have 
\begin{equation}
Q=\frac{B}{2}\pm \frac{1}{2}
\end{equation}
\noindent because $1/6=(1/2)$ $\times (1/3)=B/2$. Equation $(1)$ is in line
with the formula 
\begin{equation}
Q=I_{3}+\frac{1}{2}\left( B+S+C+B^{\ast }+T\right)
\end{equation}
where $I_{3}$ is the isospin component of the isospin, $B=1/3$ is the baryon
number and $S,C,B^{\ast },T$ denote the quark numbers for the quarks $s,c,b$
and $t$, respectively. $C$ and $T$ are equal to $1$ and $S$ and $B^{\ast }$
are equal to $-1$. The above formula (Eq. $(2)$) can also be written as \ 
\begin{equation}
Q=I_{3}+\frac{Y}{2}
\end{equation}
which is the Gell-Mann--Nishijima relation where $Y$ is the strong
hypercharge.

Since for primons each $B=1/6$ Eq. $(1)$ is not valid. Instead of it we
should have 
\begin{equation}
Q=2B\pm \frac{1}{2}
\end{equation}
because electric charges are symmetric about $1/3$ (Figure 1) since $%
1/3=2\times 1/6=2B$. This implies that for a system of primons 
\begin{equation}
Q=2B+\frac{1}{2}\left( P_{1}+P_{2}+P_{3}+P_{4}\right)
\end{equation}
\noindent where $B$ is the total baryon number, and\ $P_{1}=1$, $%
P_{2}=P_{3}=P_{4}=-1$.\ From this we note that we may divide primons into
two distinct categories and we should search for a new quantum number to
caracterize such distinction. Therefore, we have 
\[
\frac{2}{3}=2\times \left( \frac{1}{6}+\frac{1}{6}\right) +\frac{1}{2}%
(1-1)\;\;for\;u,c,t 
\]
and 
\[
-\frac{1}{3}=2\times \left( \frac{1}{6}+\frac{1}{6}\right) +\frac{1}{2}%
(-1-1)\;\;for\;d,s,b. 
\]

Because quarks $u$ and $d$ have isospin equal to $1/2$ and $-1/2$,
respectively, we are forced to make $I_{3}=\pm 1/4$ for primons $p_{1} $ and 
$p_{2}$. We will see in detail below how $I_{3}$ can be assigned to them.
And we will be able then to find that the other primons also have $I_{3}=\pm
1/4$.

Let us try to write a simple expression for the charge of primons like the
one that is used for nucleons. Following the footsteps of the strong
hypercharge we can try to make 
\begin{equation}
Q=I_{3}+\frac{Y}{2}
\end{equation}
where $Y$ is the new hypercharge and is given by 
\begin{equation}
Y=B+\Sigma
\end{equation}
where $\Sigma $ is a new quantum number to be found. Thus the formula
becomes 
\begin{equation}
Q=I_{3}+\frac{1}{2}\left( B+\Sigma _{3}\right)
\end{equation}
which is quite similar to the Gell-Mann--Nishijima formula used for quarks 
\[
Q=I_{3}+\frac{1}{2}\left( B+S\right) . 
\]

From now on instead of dealing with the new hypercharge $Y=B+\Sigma $ we
will deal directly with $\Sigma $. As was discussed above we should try $%
\Sigma _{3}=+1$ for $p_{1}$\bigskip and $\Sigma _{3}=-1$ for $p_{2}$, $p_{3}$%
, $p_{4}$. Therefore, 
\[
\frac{5}{6}=\frac{1}{4}+\frac{1}{2}\left( \frac{1}{6}+1\right) \;for\;p_{1} 
\]
and 
\[
-\frac{1}{6}=\frac{1}{4}+\frac{1}{2}\left( \frac{1}{6}-1\right) \;for\;\text{
}p_{2},p_{3},p_{4}\text{.} 
\]
As we will shortly see $p_{2},p_{3},p_{4}$ can also have $I_{3}=-1/4$. In
this case we have 
\[
-\frac{1}{6}=-\frac{1}{4}+\frac{1}{2}\left( \frac{1}{6}+0\right)
\;for\;p_{2},p_{3},p_{4}\text{.} 
\]
This means thus that $\Sigma _{3}$ can assume the values $-1$, $0$, and $+1$
and, thus, they can be considered as the projections of $\Sigma =1$. Of
course $\Sigma _{3}=0$ can also be the projection of $\Sigma =0$. Putting
all together in a table one has Table 4 below.

\begin{center}
\begin{tabular}{|c||c|c|}
\hline
&  &  \\ 
& $I_{3}$ & $\Sigma _{3}$ \\ 
&  &  \\ \hline\hline
&  &  \\ 
$p_{1}$ & $+\frac{1}{4}$ & $+1$ \\ 
&  &  \\ \hline
&  &  \\ 
$p_{j}$ & $+\frac{1}{4}$ & $-1$ \\ 
$(j=2,3,4)$ &  &  \\ \cline{2-3}
&  &  \\ 
& $-\frac{1}{4}$ & $0$ \\ 
&  &  \\ \hline
\end{tabular}
\end{center}

\bigskip

Let us now find the values of $\Sigma $ for quarks. The results are quite
impressive because they are directly linked to the Kobayashi-Maskawa matrix
and to the quark doublets

\[
\left( 
\begin{array}{c}
u \\ 
d
\end{array}
\right) ,\left( 
\begin{array}{c}
c \\ 
s
\end{array}
\right) ,\left( 
\begin{array}{c}
t \\ 
b
\end{array}
\right) 
\]

As was seen above $p_{1}$ can only have $I_{3}=1/4$, and $p_{2},p_{3},p_{4}$
can have $I_{3}=\pm 1/4$. In the case of the $u$ quark $p_{1}$ has $%
I_{3}=1/4 $ and $p_{2}$ has $I_{3}=1/4$ because the $I_{3}$ of the $u$ quark
is $1/2,$ and \ the total value of $\Sigma _{3}$ is $+1+(-1)=0$. For the $d$
quark $p_{2}$ has $I_{3}=-1/4$ and $p_{3}$ has $I_{3}=-1/4$ because the $%
I_{3}$ of the $d$ quark is $-1/2$, and the total $\Sigma _{3}$ is thus $%
0+0=0 $. \ In the case of $c$ and $t$\ quarks, since $p_{1}$ has $I_{3}=1/4$%
, $p_{2}$\ and $p_{3}$ should have $I_{3}=-1/4$ because the $I_{3}$ of both
quarks is equal to $0$. In both cases the total $\Sigma _{3}$ is equal to $%
+1 $. For quarks $s $ and $b$, their primons have to have opposite $I_{3}$
's $\ $because the $I_{3}$ of both of them is equal to $0$. And thus in both
cases the total $\Sigma _{3}$ is equal to $-1$. Hence a system of two
primons (that is, a quark) has four possible states $|\Sigma ,\Sigma _{3}>$:

\bigskip

\begin{center}
\begin{tabular}{|c|c|}
\hline
$|1,1>=c,t$ &  \\ \hline
$|1,0>=d$ & $|0,0>=u$ \\ \hline
$|1,-1>=s,b$ &  \\ \hline
\end{tabular}
\end{center}

\bigskip

The choice $|1,0>=d$, $|0,0>=u$ was made considering that the $u$ quark is
the end product of the decays of quarks which means that it should be
singled out. Making a table with the results we obtain Table 5

\begin{center}
\begin{tabular}{|c||c|c|}
\hline
&  &  \\ 
& $I_{3}$ & $\Sigma_{3}$ \\ 
&  &  \\ \hline\hline
&  &  \\ 
$c,t$ & $0$ & $+1$ \\ 
&  &  \\ \hline
&  &  \\ 
$u$ & $+\frac{1}{2}$ & $0$ \\ 
&  &  \\ \hline
&  &  \\ 
$d$ & $-\frac{1}{2}$ & $0$ \\ 
&  &  \\ \hline
&  &  \\ 
$s,b$ & $0$ & $-1$ \\ 
&  &  \\ \hline
\end{tabular}
\end{center}

\bigskip

Making a diagram with this table we obtain Figure 2 
\[
\left( 
\begin{array}{c}
|0,0> \\ 
|1,0>
\end{array}
\right) ,\left( 
\begin{array}{c}
|1,1> \\ 
|1,-1>
\end{array}
\right) ,\left( 
\begin{array}{c}
|1,1> \\ 
|1,-1>
\end{array}
\right) 
\]
which should be compared to 
\[
\left( 
\begin{array}{c}
u \\ 
d
\end{array}
\right) ,\left( 
\begin{array}{c}
c \\ 
s
\end{array}
\right) ,\left( 
\begin{array}{c}
t \\ 
b
\end{array}
\right) 
\]

Comparing Figure 2 (or Table 5) with the Kobayashi-Maskawa matrix$^{9,10}$
we note that the elements of the matrix $|U_{cs}|$ and $|U_{tb}|$ (which are
the largest ones are almost equal to $1$) satisfy the selection rule $\Delta
\Sigma _{3}=-2$, $\Sigma =0$, and the other large element $|U_{ud}|$ (which
is also close to $1$) satisfies the selection rule, $\Delta \Sigma =-1$, $%
\Sigma _{3}=0$. The second largest elements $|U_{cd}|$ ($=0.24$) and $%
|U_{us}|$ ($=0.23$) obey, respectively, the selection rules $\Delta \Sigma
_{3}=-1$, and $\Delta \Sigma _{3}=+1,$ $\Delta \Sigma =-1$ . The null
elements (or almost null) $|U_{ts}|$ and $|U_{cb}|$ can also be understood
according to the above scheme if we also take into account the three quark
doublets. According to the scheme flavor changing neutral currents are
forbidden because in such a case $\Delta \Sigma _{3}=0$.\ Taking a glance at
the above diagram we can understand why $|U_{cs}|\approx |U_{tb}|$ and why $%
|U_{td}|\approx $ $|U_{ub}|$. {\bf Another very important conclusion is that
the }$c${\bf \ and }$t${\bf \ quarks are heavier versions of the }$s${\bf \
and }$b${\bf \ quarks, respectively.} {\bf Thus, we can say that the general
rule behind the weak decay of quarks is: a quark decays to diminish either
its }$\Sigma ${\bf \ or its }$\Sigma _{3}${\bf .} When $\Sigma =1$ one way
is to go from $\Sigma _{3}=+1$ to $\Sigma _{3}=-1,0$, and another way is to
go from $\Sigma _{3}=-1$ to $\Sigma _{3}=0$ (but between \ $\Sigma =1$ and $%
\Sigma =0$), and\ when $\Sigma _{3}=0$ the only way left is to go from $%
\Sigma =1$ to $\Sigma =0$. If we represent the values of $\Sigma _{3}$ by
arrows as we do with spin we find a complete similarity with the case of $%
S=1 $ for two $1/2$ spin particles or with $I=1$)(isospin) of the
nucleon-nucleon system. That is, 
\[
\begin{array}{c}
c=p_{1}\uparrow p_{3}\uparrow ;\;t=p_{1}\uparrow p_{4}\uparrow ;\;|1,1> \\ 
d=\frac{1}{\sqrt{2}}\left( p_{2}\uparrow p_{3}\downarrow +p_{2}\downarrow
p_{3}\uparrow \right) ;\;|1,0> \\ 
s=p_{2}\downarrow p_{4}\downarrow ;\;b=p_{3}\downarrow p_{4}\downarrow
;\;|1,-1>
\end{array}
\]
and 
\[
u=\frac{1}{\sqrt{2}}\left( p_{1}\uparrow p_{2}\downarrow -p_{1}\downarrow
p_{2}\uparrow \right) ;\;|0,0> 
\]
and thus there is an $SU(2)$ related to $\Sigma $.\ This means that the
variation of $\Sigma $ should be related to the variation of weak isospin.
One clearly notices that both $t$ and $c$ have $p_{1}$, and thus, the mass
difference between them is directly linked to how $p_{3}$ and $p_{4}$\
interact with $p_{1}$ because $t$ and $c$ have the same $\Sigma $ and the
same $\Sigma _{3}$. The same happens between quarks $s$ and $b$ which have $%
p_{4}$ in common. Since a given primon takes part in heavy and in light
quarks as well ($p_{1}$, for example), primons probably have the same mass
which should be a light mass. More on this we will see below. Another
important point to consider is that by means of $\Sigma $ and $\Sigma _{3}$
we can understand why there are quark weak decays that conserve strong
isospin ($t\longrightarrow b$, $c\longrightarrow s$).

This new $SU(2)$ is in complete agreement with weak isospin and, thus the
Weinberg-Salam model (applied to quarks) does not need any deep modification
since the symmetry continues to be the same. But by means of the new $SU(2)$
we can begin to understand what is behind the Kobaiashi-Maskawa matrix: 
{\large the composition of quarks.}

\noindent {\large Towards a New Chromodynamics (Superchromodynamics(SQCD))}

As we saw in the sections above we need four primons and three supercolors
to generate quarks in the three \ \ colors. This means that in terms of
flavors primons can be represented by the Dirac spinors 
\[
\Psi _{1}=\left( 
\begin{array}{c}
1 \\ 
0 \\ 
0 \\ 
0
\end{array}
\right) ,\;\;\Psi _{2}=\left( 
\begin{array}{c}
0 \\ 
1 \\ 
0 \\ 
0
\end{array}
\right) ,\;\;\Psi _{3}=\left( 
\begin{array}{c}
0 \\ 
0 \\ 
1 \\ 
0
\end{array}
\right) ,\;\;\Psi _{4}=\left( 
\begin{array}{c}
0 \\ 
0 \\ 
0 \\ 
1
\end{array}
\right) . 
\]
\newline
As to supercolors we may represent them by the three-element columns 
\[
sc_{\alpha }=\left( 
\begin{array}{c}
1 \\ 
0 \\ 
0
\end{array}
\right) ,\;\;sc_{\beta }=\left( 
\begin{array}{c}
0 \\ 
1 \\ 
0
\end{array}
\right) ,\;\;sc_{\gamma }=\left( 
\begin{array}{c}
0 \\ 
0 \\ 
1
\end{array}
\right) . 
\]

\bigskip\ \noindent The supercolor generators are the three-dimensional
generators of SU(2) 
\[
{\Theta }^{1}=\left( 
\begin{array}{ccc}
0 & 0 & 0 \\ 
0 & 0 & -i \\ 
0 & i & 0
\end{array}
\right) ,\;\;{\Theta }^{2}=\left( 
\begin{array}{ccc}
0 & 0 & i \\ 
0 & 0 & 0 \\ 
-i & 0 & 0
\end{array}
\right) ,\;\;{\Theta }^{3}=\left( 
\begin{array}{ccc}
0 & -i & 0 \\ 
i & 0 & 0 \\ 
0 & 0 & 0
\end{array}
\right) , 
\]
which are three of the eight generators of $SU(3)$, and obey the relations 
\begin{equation}
\lbrack \Theta ^{j},\Theta ^{k}]=i\varepsilon ^{jkl}\Theta ^{l}.
\end{equation}

\noindent Let us call them supergluons. Such as gluons supergluons are
vectorial ($S=1$) and also massless. According to the ideas above developed
the combinations of equal supercolors do not produce a color. That is
exactly what we have: 
\[
{sc_{\alpha }^{\dagger }\Theta }{^{j}}sc_{\alpha }={sc_{\beta }^{\dagger
}\Theta ^{j}}sc_{\beta }={sc_{\gamma }^{\dagger }\Theta ^{j}}sc_{\gamma }=0, 
\]
where $j=1,2,3$. With different supercolors we have the sums: \newline
\[
\sum_{j=1}^{3}{sc_{\alpha }^{\dagger }\Theta }^{j}sc_{\beta
}=-i,\;\;\sum_{j=1}^{3}{sc_{\alpha }^{\dagger }\Theta }^{j}sc_{\gamma
}=i,\;\;\sum_{j=1}^{3}{sc_{\beta }^{\dagger }\Theta }^{j}sc_{\gamma }=-i. 
\]
\newline
\noindent Therefore, the substructure of $SU(3)$(color) is $SU(2)$%
(supercolor).

\noindent {\large The Lagrangian of Quantum Superchromodynamics }

\bigskip Following the footsteps of QCD we can propose that the free
Lagrangian for primons is 
\begin{equation}
L=i{\hbar }c\overline{\Psi }\gamma ^{\mu }\partial _{\mu }\Psi -mc^{2}%
\overline{\Psi }\Psi
\end{equation}
\noindent in which $\Psi \;$ is the column 
\begin{equation}
\Psi =\left( 
\begin{array}{c}
\Psi _{\alpha } \\ 
\Psi _{\beta } \\ 
\Psi {\gamma }
\end{array}
\right)
\end{equation}

\noindent and $\Psi _{i}$ is a four-component Dirac spinor. \noindent In the
same way as is done in QCD we can construct the gauge invariant (under
supercolor SU(2)) QSCD Lagrangian 
\begin{equation}
L=i{\hbar }c\overline{\Psi }\gamma ^{\mu }\partial _{\mu }\Psi -mc^{2}%
\overline{\Psi }\Psi -\frac{1}{16{\pi }}\Gamma ^{\mu {\nu }}\Gamma _{\mu {%
\nu }}-g_{sc}\overline{\Psi }\gamma ^{\mu }{\Theta }{\Psi }A_{\mu }
\end{equation}
\noindent in which $g_{sc}$ is the supercolor coupling constant, and $\Gamma
^{\mu \nu }\;$ are the supergluon fields. The above Lagrangian should hold
for each primon because there are four different mass terms. That is, there
are four different Lagrangians.

Since primons are almost massless the Lagrangian (for each primon) can be
written as 
\begin{equation}
L=i{\hbar }c\overline{\Psi }\gamma ^{\mu }\partial _{\mu }\Psi -\frac{1}{%
16\pi }\Sigma ^{\mu \nu }\Sigma _{\mu \nu }-g_{sc}\overline{\Psi }\gamma
^{\mu }{\Theta }{\Psi }A_{\mu }.
\end{equation}
\noindent The last two Lagrangians above are invariant under local SU(2)
gauge transformations and describe the interaction of each primon (that is,
each flavor) with the three massless vector fields (supergluons). The Dirac
fields make the three supercolor currents 
\begin{equation}
I^{\zeta }=cg_{sc}\bar{\Psi}\gamma ^{\zeta }{\Theta }\Psi
\end{equation}
\noindent which are the sources of the supercolor fields.

\bigskip

\noindent {\large The Masses of Primons}

The magnetic moments of primons should be given by 
\[
\begin{array}{c}
\mu _{1}=\frac{5}{6}\frac{e}{2m_{1}}\;for\;p_{1} \\ 
\mu _{2}=-\frac{1}{6}\frac{e}{2m_{2}}\;for\;p_{2},p_{3},p_{4}
\end{array}
\]
and hence 
\[
\mu _{2}=-\frac{5m_{2}}{m_{1}}\mu _{1} 
\]
and 
\[
\mu _{3}=\frac{m_{2}}{m_{3}}\mu _{2}. 
\]
Considering that the spin content of quarks should be the same we have 
\[
\frac{\mu _{u}}{\mu _{d}}=\frac{\mu _{1}+\mu _{2}}{\mu _{2}+\mu _{3}} 
\]
and since $\mu _{u}/\mu _{d}=-2$ and using the above relations we obtain 
\[
-2=-\frac{5m_{2}}{m_{1}}\left( \frac{1-\frac{m_{1}}{5m_{2}}}{1+\frac{m_{2}}{%
m_{3}}}\right) . 
\]
Making $m_{3}=fm_{2}$ and solving for the ratio $m_{1}/m_{2}$ we arrive at 
\[
\frac{m_{1}}{m_{2}}=\frac{5}{3+\frac{2}{f}} 
\]
and since the mass of $u(p_{1}p_{2})$ and $d(p_{2}p_{3})$ are about the same
it is reasonable to suppose that $m_{3}\approx m_{1}$ and thus 
\[
f\approx \frac{5}{3+\frac{2}{f}} 
\]
which yields $f\approx 1$ and hence $m_{3}\approx m_{2}$. Then it is
reasonable to assume that primons have about the same mass.

\noindent {\large The Origin of Quark Mass}

In order to have very light primons we can consider that every pair of
primons of a quark are bound by means of a very strong spring. Thus the mass
of each quark should be equal to 
\begin{equation}
m_{q}c^{2}\approx \frac{{\hbar \omega }}{2}=\frac{{\hbar }}{2}\sqrt{\frac{k}{%
\mu _{p}}}
\end{equation}

\noindent in which $\mu _{p}$ is the reduced mass of the pair of primons and 
$k$ is the effective constant of the spring between them. It is worth
mentioning that a quite similar idea is used for explaining quark
confinement and based on it a term $Kr$ is introduced in the effective
potential. For the $u$ quark, for example, we have $m_{u}c^{2}\approx 0.3$%
GeV. On the other hand if we consider a harmonic potential we have 
\begin{equation}
m_{q}c^{2}\approx \frac{1}{2}k_{u}(R_{q})^{2}
\end{equation}
where $R_{q}$ is the size of the quark $q$. For $u$ we obtain $k_{u}\approx
10^{20}J/m^{2}\approx 2GeV/fm^{2}.$ Using this figure above we obtain $\mu
_{p}\approx 10^{-28}kg$ which is about the proton mass. Therefore, in order
to have light primons the effective well has to have a larger dependence
with the distance between the two primons. Considering that the potential is
symmetrical about the equilibrium position we may try to use the potential 
\begin{equation}
V(x)=\alpha _{u}x^{4}.
\end{equation}
The energy levels of the potential $V(x)=\alpha x^{\upsilon }$ are given by$%
^{24}$%
\begin{equation}
E_{n}=\left[ \sqrt{\frac{\pi }{2\mu }}\nu \hbar a^{1/\nu }\frac{\Gamma (%
\frac{3}{2}+\frac{1}{\nu })}{\Gamma (\frac{1}{\nu })}\right] ^{2\nu /(2+\nu
)}(n+\frac{1}{2})^{2\nu /(2+\nu )}.
\end{equation}
Thus, for $\nu =4$ and $n=0$ we have

\begin{equation}
E_{0}=\left[ \sqrt{\frac{\pi }{2\mu _{p}}}4\hbar a_{u}^{1/4}\frac{\Gamma (%
\frac{3}{2}+\frac{1}{4})}{\Gamma (\frac{1}{4})}\right] ^{4/3}(0+\frac{1}{2}%
)^{4/3}
\end{equation}

\noindent and then we obtain (making $E_{0}=m_{q}c^{2}$) 
\begin{equation}
\mu _{p}\sim 0.25\hbar ^{2}\sqrt{\frac{a_{u}}{(m_{q}c^{2})^{3}}}
\end{equation}
\noindent which can be extremely light depending on the value of $a_{u}.$ \
The above figure should be taken with caution because it is a result of
nonrelativistic quantum mechanics but it does not change the fact that
primons may have a very small mass.

Thus, if primons interact via a very strong potential such as $V(x)=\alpha
_{u}x^{4}$ they can be extremely light fermions. We can then propose a more
general effective potential of the form $V(x)=\frac{1}{4}\alpha _{u}x^{4}-%
\frac{1}{2}k_{u}x^{2}$ where the last term is chosen negative. Generalizing
the coordinate $x$ we can consider the \ ``potential energy '' 
\begin{equation}
V_{0}(\phi )=\frac{1}{4}\lambda ^{2}\phi ^{4}-\frac{1}{2}\mu ^{2}\phi ^{2}
\end{equation}
where $\phi $ is a field related to the presence of the two primons (or of
other primons of the same baryon) and is the scalar interaction between
them, and $\mu $ and $\lambda $ are real constants. Hence we can propose the
Lagrangian 
\begin{equation}
{\cal L}_{0}{\cal =}\frac{1}{2}(\partial _{\nu }\phi )(\partial ^{\nu }\phi
)+\frac{1}{2}\mu ^{2}\phi ^{2}-\frac{1}{4}\lambda ^{2}\phi ^{4}
\end{equation}
between the two primons of a quark. Since a quark only exists by means of
the combination of the two primons we may consider that its initial mass is
very small. The above Lagrangian is symmetric in $\phi $ but let us recall
that primons can interact by other means, electromagnetically, for example.
Therefore, we can make the transformation $\phi \longrightarrow \phi +\eta
_{ev}$ where $\eta _{ev}$ is a deviation caused by the electromagnetic field
and vacuum. The new potential energy up to second power in $\eta _{ev}$ is 
\begin{equation}
V(\phi ,\eta _{ev})=V_{0}-\mu ^{2}\phi \eta _{ev}-\frac{1}{2}\mu ^{2}\eta
_{ev}^{2}+\lambda ^{2}\phi ^{3}\eta _{ev}+\frac{3}{2}\lambda ^{2}\phi
^{2}\eta _{ev}^{2}.
\end{equation}

\noindent \bigskip\ $V(\phi ,\eta _{ev})$ has a minimum at 
\begin{equation}
\eta _{ev}(\phi )=\frac{-\mu ^{2}\phi +\lambda ^{2}\phi ^{3}}{\mu
^{2}-3\lambda ^{2}\phi ^{2}}.
\end{equation}

\noindent Since $\eta _{ev}$ is small let us make $\mu ^{2}\phi -\lambda
^{2}\phi ^{3}=\delta $ (a small quantity). Then we can make $\phi \approx
\pm \frac{\mu }{\lambda }+\epsilon $ and obtain $\epsilon \approx -\frac{%
\delta }{2\mu ^{2}}$ and thus \ 
\begin{equation}
\phi \approx \pm \frac{\mu }{\lambda }-\frac{\delta }{2\mu ^{2}}
\end{equation}
\noindent and the symmetry has disappeared. But it is not spontaneously
broken, it is broken by the perturbation $\eta _{ev}$ which may be linked to
charge. Substituting the above value of $\phi $ into Eq 20 we have

\begin{equation}
U(\eta _{ev})=V(\phi ,\eta _{ev})-V_{0}\approx \mu ^{2}\eta _{ev}^{2}
\end{equation}
and the approximate Lagrangian is 
\begin{equation}
{\cal L=}\frac{1}{2}(\partial _{\nu }\eta _{ev})(\partial ^{\nu }\eta
_{ev})-\mu ^{2}\eta _{ev}^{2}
\end{equation}
which is a Klein-Gordon Lagrangian with mass 
\begin{equation}
m=\sqrt{2}\mu \hbar /c
\end{equation}
which may be an effective mass. Taking a look at Table 3 we observe that we
need three scalar bosons, $\eta _{ev}^{+}$, $\eta _{ev}^{-}$ and $\eta
_{ev}^{0}$. The first and second \ particles are exchanged between the
primons of the quarks $p_{1}p_{2}(u)$, $p_{1}p_{3}(c)$, and $p_{1}p_{4}(t)$,
and the neutral boson is exchanged between the primons of the quarks $%
p_{2}p_{3}(d)$, $p_{2}p_{4}(s)$, and $p_{3}p_{4}(b)$. Therefore, three Higgs
bosons produce the masses of quarks.

It is quite interesting that we should have a triplet of scalar bosons. And
we notice immediately a very important trend: The charged bosons produce
masses larger\ than those produced by the neutral boson, considering the
quark generations

\begin{center}
\bigskip $\left( 
\begin{array}{c}
u \\ 
d
\end{array}
\right) ,\left( 
\begin{array}{c}
c \\ 
s
\end{array}
\right) ,\left( 
\begin{array}{c}
t \\ 
b
\end{array}
\right) .$
\end{center}

{\large Therefore, the origin of mass in quarks is also linked to the values
of their charges.}

\noindent This is summarized below in Table 6.

\bigskip \pagebreak 

\begin{center}
\begin{tabular}{||ccc||ccc|ccc|ccc||}
\hline\hline
&  &  &  &  &  &  &  &  &  &  &  \\ 
& Quark &  &  & Mass(GeV) &  &  & Charge &  &  & Mass Generator &  \\ 
&  &  &  &  &  &  &  &  &  & (Higgs Bosons) &  \\ \hline\hline
&  &  &  &  &  &  &  &  &  &  &  \\ 
& $u(p_{1}p_{2})$ &  &  & $0.3$ &  &  & $+\frac{2}{3}$ &  &  & $\eta
_{ev}^{+}$, $\eta _{ev}^{-}$ &  \\ 
&  &  &  &  &  &  &  &  &  &  &  \\ \hline
&  &  &  &  &  &  &  &  &  &  &  \\ 
& $c(p_{1}p_{3})$ &  &  & $1.5$ &  &  & $+\frac{2}{3}$ &  &  & $\eta
_{ev}^{+}$, $\eta _{ev}^{-}$ &  \\ 
&  &  &  &  &  &  &  &  &  &  &  \\ \hline
&  &  &  &  &  &  &  &  &  &  &  \\ 
& $t(p_{1}p_{4})$ &  &  & $170$ &  &  & $+\frac{2}{3}$ &  &  & $\eta
_{ev}^{+}$, $\eta _{ev}^{-}$ &  \\ 
&  &  &  &  &  &  &  &  &  &  &  \\ \hline
&  &  &  &  &  &  &  &  &  &  &  \\ 
& $d(p_{2}p_{3})$ &  &  & $\gtrapprox 0.3$ &  &  & $-\frac{1}{3}$ &  &  & $%
\eta _{ev}^{0}$ &  \\ 
&  &  &  &  &  &  &  &  &  &  &  \\ \hline
&  &  &  &  &  &  &  &  &  &  &  \\ 
& $s(p_{2}p_{4})$ &  &  & $0.5$ &  &  & $-\frac{1}{3}$ &  &  & $\eta
_{ev}^{0}$ &  \\ 
&  &  &  &  &  &  &  &  &  &  &  \\ \hline
&  &  &  &  &  &  &  &  &  &  &  \\ 
& $b(p_{3}p_{4})$ &  &  & $4.5$ &  &  & $-\frac{1}{3}$ &  &  & $\eta
_{ev}^{0}$ &  \\ 
&  &  &  &  &  &  &  &  &  &  &  \\ \hline\hline
\end{tabular}
\end{center}

\vspace*{0.2in}

\begin{center}
\parbox{4in}
{Table 6. The masses of quarks and their generators. As is known the mass of 
the $d$ quark is slightly larger than that of the $u$ quark. There is a clear
division between the three first quarks and the other three quarks. The quarks
generated by the charged bosons have larger masses and larger charges and 
those generated by $\eta_{ev}^{0}$ have smaller masses and smaller charges}.
\end{center}

\bigskip

\noindent \noindent {\large The Arrangement of Primons in Baryons}

A possible way for the arrangement of primons in baryons is discussed in
references 5, 6, 7 and 8. Such arrangement agrees well with the electric
charge distribution of the nucleons, with the nucleon dipole moments, with
the stability of the alpha particle and of the deuteron, with the absence of
nuclides with A=5, and with the enormous instability of $Be^{8}$.

\noindent {\large The Spin of Primons (and Confinement)}

The spin of primons is a great puzzle in the same way as the spin of a quark
is too since as is well known only half of the spins of quarks contribute to
the total spin of a baryon as found by the experiments. Actually, the spin
puzzle of baryons may be linked to the spin of primons. Since they are
elementary fermions \ we expect them to be half spin fermions. And thus it
is not an easy task to divise a way of \ making two spin half fermions to
compose another spin half fermion since the space of spin {\bf is not the
normal three-dimensional space}. That is why the scheme divised in
references 6 and 7 is not very good and may be wrong. When we make the
addition of two different angular momenta we use the commutation relation 
\begin{equation}
\left[ \overrightarrow{S_{1}},\overrightarrow{S_{2}}\right] =0
\end{equation}
which means that their degrees of freedom are independent. Maybe in the case
of primons they are not independent since each set of primons is a baryon,
that is, in the case of spin the problem to be solved is the problem of
composing the total spin of a set of six particles subject to the rigid
condition of obtaining a baryon in the end. This means that primons only
exist forming baryons (and mesons) and thus free primons are not possible.
Therefore, they are always confined.

\noindent \qquad This article intends to be just a preliminary work on quark
composition. It appears to me that such composition is firmly established
(indirectly) by the Kobaiashi-Maskawa matrix. Much more work should be and
will be developed on this subject.

\bigskip \bigskip 

\noindent {\Large References}

\noindent 1. E.J. Eichten, K.D. Lane, and M.E. Peskin, Phys. Rev. Lett. 50,
811 (1983). \newline
\noindent 2. K. Hagiwara, S. Komamiya, and D. Zeppenfeld, Z. Phys. C29, 115
(1985). \newline
\noindent 3. N. Cabibbo, L. Maiani, and Y. Srivastava, Phys. Lett. 139, 459
(1984). \newline
\noindent 4. H. Fritzsch, in {\it Proceedings of the twenty-second Course of
the International School of Subnuclear\ Physics}, 1984, ed. by A. Zichichi
(Plenum Press, New York, 1988). \newline
\noindent 5.\noindent\ M. E. de Souza, in {\it Proceedings of the XII
Brazilian National Meeting of the Physics of Particles and\ Fields},
Caxambu, Minas Gerais, Brazil, September 18-22, 1991.\newline
\noindent 6. M.E. de Souza, in {\it The Six Fundamental Forces of Nature},
Universidade Federal de Sergipe, S\~{a}o Crist\'{o}v\~{a}o, Sergipe, Brazil,
February 1994. \newline
\noindent 7. M.E. de Souza, in {\it The General Structure of Matter},
Universidade Federal de Sergipe, S\~{a}o Crist\'{o}v\~{a}o, Sergipe, Brazil,
July 2001. \newline
\noindent 8. M.E. de Souza, hep/ph 0207301.

\noindent 9. F. Halzen and A.D. Martin, in {\it Quarks \& Leptons: An
Introductory Course in Modern Particle Physics, }John Wiley \& Sons, New
York, 1984, p. 286.

\noindent 10. P. Renton, in {\it Electroweak Interactions, An Introduction
to the Physics of Quarks \& Leptons}, Cmbridge University Press, Cambridge,
1990, pp. 428,429.

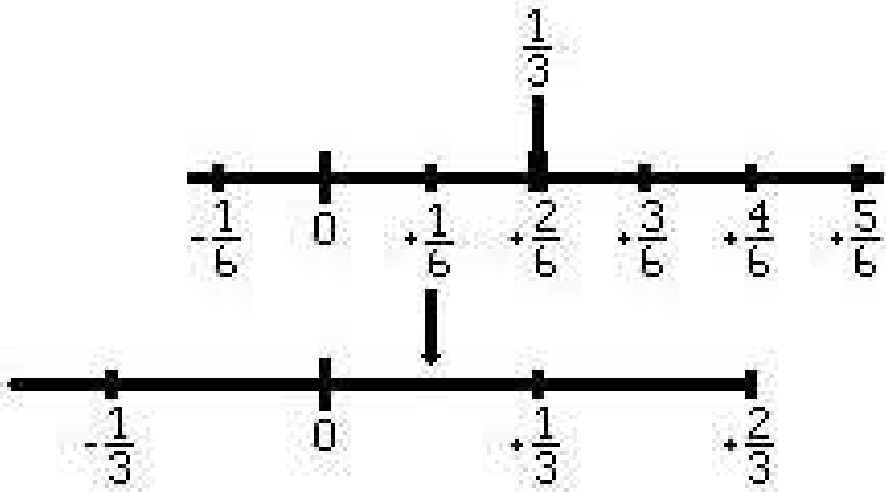
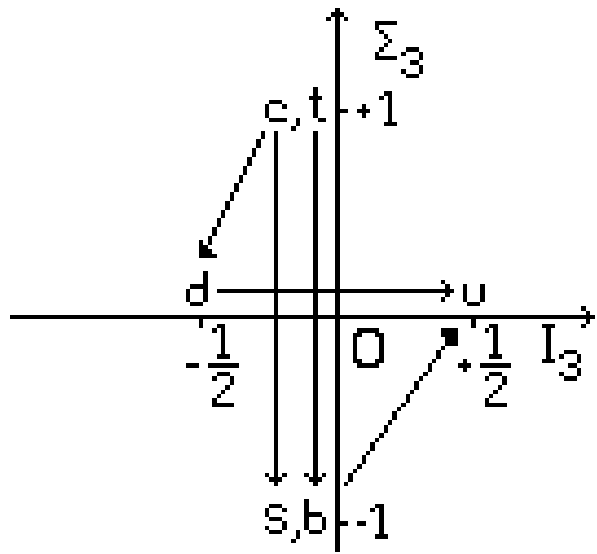
\end{document}